\documentclass[aps,prl,reprint,superscriptaddress,amsmath,amssymb]{revtex4-2}

\usepackage{graphicx}
\usepackage{bm}
\usepackage{tikz}
\usepackage{pgfplots}
\pgfplotsset{compat=1.17}
\usepackage[colorlinks=true,linkcolor=blue,citecolor=blue,urlcolor=blue]{hyperref}
\usepackage{mathrsfs}
\usepackage{braket}
\usepackage[normalem]{ulem}
\definecolor{c2}{RGB}{255,127,14}
\definecolor{c3}{RGB}{44,160,44}
\definecolor{cplus}{RGB}{31,119,180}
\definecolor{cminus}{RGB}{214,39,40}

\newcommand{\nh}{\hat n}

\newcommand{\cP}{\mathcal{P}}
\newcommand{\cD}{\mathcal{D}}
\newcommand{\cR}{\mathcal{R}}

\newcommand{\Scri}{\mathscr{I}}
\newcommand{\bz}{\bar z}
\newcommand{\Li}{\operatorname{Li}_2}
\newcommand{\arctanhh}{\operatorname{arctanh}}

\begin{document}

\title{Electromagnetic Memory as a Collider Observable}

\author{Felipe Diaz}
\email{felipe.diaz@uss.cl}
\affiliation{Facultad de Ingenier\'ia, Universidad San Sebasti\'an, Santiago 8420524, Chile}
\author{Hern\'an A. Gonz\'alez}
\email{hernan.gonzalez@uss.cl}
\affiliation{Facultad de Ingenier\'ia, Universidad San Sebasti\'an, Santiago 8420524, Chile}
\author{Oriana Labrin}
\email{olabrinz@correo.uss.cl}
\affiliation{Facultad de Ingenier\'ia, Universidad San Sebasti\'an, Santiago 8420524, Chile}
\author{Jakob Salzer}
\email{jakob.salzer@tuwien.ac.at}
\affiliation{Institute for Theoretical Physics, TU Wien, Wiedner Hauptstrasse 8--10, 1040 Vienna, Austria}
\affiliation{Universit\'e Libre de Bruxelles and International Solvay Institutes, 
ULB-Campus Plaine CP231, 1050 Brussels, Belgium}

\date{\today}

\begin{abstract}
Electromagnetic memory and charge correlators are measured by different detectors, sensitive to soft radiation and to hard charged tracks. We show that the Ward identity of large gauge symmetry relates these quantities as inclusive observables. For lepton pairs produced by an electromagnetic current, we compute all massless multipoint memory correlators and the massive two-point function, whose coincident limit gives the soft photon number. The correlators admit a celestial free-boson representation and remain finite at first radiative order, where recoil generates back-to-back Sudakov logarithms.
\end{abstract}

\maketitle
 
\emph{Introduction.}---Every process that produces charged particles leaves a permanent imprint on the electromagnetic field at infinity. This electromagnetic memory~\cite{Bieri:2013hqa,Pasterski:2015zua} connects classical radiation and the soft photon theorem~\cite{Low:1958sn, Weinberg:1965nx}. Its direct detection remains challenging~\cite{Bieri:2023eme,Zosso:2025eme}.   However, the collision of charged particles offers another way to access it, through the correlations measured at colliders. In quantum electrodynamics (QED), the soft photon theorem can be formulated as a Ward identity for a large gauge transformation~\cite{He:2014cra,Kapec:2015ena,Strominger:2017zoo}. The generator of this symmetry contains a hard contribution, determined by the distribution of electric charges, and a soft contribution obtained from the soft photon zero mode \cite{Ashtekar:1987tt, Donnay:2018neh}. This gives an operational meaning to the soft photon insertion as an insertion of soft charges, that can be related to charge detector operators~\cite{Hofman:2008ar, Lee:2023npz, Monni:2025zyv}. This makes memory an observable that is accessible in collider experiments.

Memory detectors and their relation to asymptotic charges have recently been developed~\cite{Hu:2023geb,Caron-Huot:2023vxl,Gonzalez:2025ene,Oertel:2026wsm,Oertel:2026oqv}. The in-in formulation also relates gravitational memory correlations to energy detectors~\cite{Moult:2025njc,Herrmann:2026kue}.  Electrodynamics offers a particularly attractive setting for these developments, since both the charged-particle distributions and the associated soft radiation can be potentially measured in collider experiments.

In this Letter, we show that the Ward identity of  large gauge symmetry, usually formulated in the context of scattering amplitudes, can be applied to inclusive observables of charged particles. It bridges two sets of observables, namely charge correlations extracted from the hard final states and memory correlators built from the angular distribution of soft radiation. 
Both are reconstructed event by event and measured by accumulating events in the same spirit as energy correlators~\cite{Basham:1978zq,Hofman:2008ar,Kravchuk:2018htv,Moult:2025nhu}. As a result, asymptotic symmetries become a predictive tool in collider physics. 

\begin{figure}[t]
\centering
\begin{tikzpicture}[scale=1.05,>=stealth]
  \shade[inner color=white,outer color=gray!22] (0,0) circle (2);
  \draw[thick] (0,0) circle (2);
  \draw[gray!70] (-2,0) arc (180:360:2 and 0.5);
  \draw[gray!70,dashed] (2,0) arc (0:180:2 and 0.5); \node[gray!50!black] at (128:2.38) {${\mathscr{I}}^+$};
  \draw[line width=1.6pt,cplus,->] (32:0.30) -- (32:1.96);
  \draw[line width=1.6pt,cminus,->] (212:0.30) -- (212:1.96);
  \fill[cplus] (32:2) circle (2.2pt);
  \fill[cminus] (212:2) circle (2.2pt);
  \node[cplus,anchor=south west] at (32:2.05) {$+\,,\ \nh_a$};
  \node[cminus,anchor=north east] at (212:2.05) {$-\,,\ -\nh_a$};
  \shade[ball color=gray!55] (0,0) circle (0.27);
  \node at (0,0) {\small $J^\mu$};
  \node[anchor=north] at (0,-0.3) {\footnotesize $q$};
  \foreach \a/\l in {95/1,155/2,318/3}{
    \draw[line width=3pt,black!80] (\a-4:2) arc (\a-4:\a+4:2);
    \node at (\a:2.42) {$\nh_{\l}$};
  }
  \draw[->,thick,black!60] (95:1.72) -- ++(185:0.35);
  \draw[->,thick,black!60] (155:1.72) -- ++(255:0.3);
  \draw[->,thick,black!60] (318:1.72) -- ++(48:0.32);
\end{tikzpicture}
\caption{At tree level, the unpolarized current $J^\mu$ produces a back-to-back pair of massless leptons, which pierce the celestial sphere at $\pm\nh_a$. In each event, the Ward identity predicts the memory at the detector directions $\nh_1,\nh_2,\dots$ (arrows) from the charged tracks through Eq.~\eqref{eq:memory}. Memory correlators average products of these values over the orientation of the pair.}
\label{fig:setup}
\end{figure}
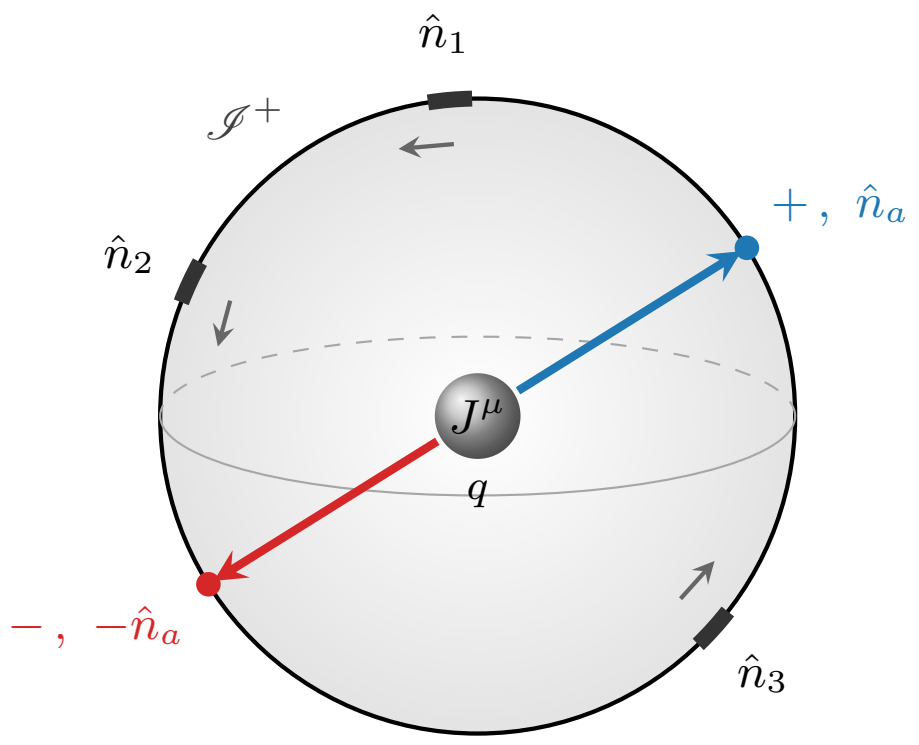

More precisely, the in-in Ward identity for QED enables us to express memory insertions produced by scattering as an integral transform of the charge-weighted cross sections. We calculate these observables for a lepton pair produced by the electromagnetic current in QED (Fig.~\ref{fig:setup}). Averaging over the current polarization makes the pair axis isotropic in the source rest frame, which allows analytic expressions for all massless multipoint functions and for the massive two-point function. Each two-body event has the same memory pattern with a different orientation. The one-point function therefore vanishes, while correlations of an even number of insertions display the angular radiation of the charged pair.

Crucially, memory correlators are finite at next-to-leading order (NLO). A real photon tilts the charged pair away from the back-to-back configuration, but the normalized correlators remain finite. The defining property of collider observables such as energy correlators is infrared safety, and at this order memory correlators share a similar feature. The recoil only leaves Sudakov logarithms near the antipode.

These results have a natural realization on the celestial sphere. The memory is described by a free boson, charged particles act as vertex operators, and the pole of the memory at each charge comes from the operator product expansion with a two-dimensional current. Recoil arises from a smeared definition of the tree-level generating functional.

More broadly, memory correlators control the infrared behavior of photon number correlators. The photon number operator develops an infrared pole whose residue is determined by the product of two memory operators~\cite{Gonzalez:2025ene}. Memory correlators at coincident points then constrain the dependence of photon number correlators on the soft energy cutoff, providing an independent means of probing electromagnetic memory at colliders.

\vspace{0.3cm}

\emph{Memory from charged particles.} --- When projected to future null infinity $\Scri^+$ and integrated over retarded time $u=t-r$, the  Maxwell equations $\nabla_\mu F^{\mu\nu}=-e^2j^\nu$ with $\alpha=e^2/4\pi$ yield the constraint~\cite{He:2014cra}
\begin{equation}
  \label{eq:constraint}
D^A\Delta A_A=-4\pi\alpha\,q(\nh)+A_u^{(1)}|_{u=+\infty}-A_u^{(1)}|_{u=-\infty}\,
\end{equation}
 with $q(\nh)$ the electric charge density of the source $j^\mu$ and $A_A$ and $A^{(1)}_u$ denoting the leading components of the gauge field. Assuming the absence of radial magnetic fields at the boundaries of $\Scri^+$, this relates the change of the leading angular gauge field, i.e., \emph{the electric memory}, to the charge of massless (massive) charged matter deposited at future null (timelike) infinity and the value of the leading radial electric field at the past boundary of $\Scri^+$. For massless particles crossing the celestial sphere at stereographic coordinates $z_k$ the constraint determines
\begin{equation}
\Delta A_z(z)=-\alpha\sum_k\frac{q_k}{z-z_k}\,,
\label{eq:memory}
\end{equation} 
where the charges are measured in units of the positron charge and the memory operator is evaluated in the direction $\hat{n}(z)=\hat{n}_z$ on the celestial sphere. We have set the last two terms in \eqref{eq:constraint}  to zero. In this derivation, we have assumed vanishing incoming radiation.

More generally, $F^{(2)}_{ur}|_{u=-\infty}=A_u^{(1)}|_{u=-\infty}$ in radial gauge, obeys matching conditions that relate it to the analogous field at the future boundary of past null infinity~\cite{He:2014cra} and spatial infinity~\cite{Fuentealba:2025ekj}. This field thus defines an angle-dependent $\textrm{U}(1)$-charge when integrated against an arbitrary function on the sphere, that is conserved between past and future null infinity, $Q^-=Q^+$. As evident from~\eqref{eq:constraint}, the charge on future null infinity $Q^+$ splits into a soft piece $Q^+_S$ whose integrand is given by $e^{-2}D^A\Delta A_A$ and a hard piece $Q^+_H$ counting the outgoing massless and massive charged matter.

This conservation law can be used to compute correlators of $Q^+_S$, and thus of the memory, in terms of correlators of $Q^+_H$, i.e., massless (massive) charge correlators. Consider a state $\mathcal{O}(q)\ket{0}$ with fixed momentum $q$ created by a neutral local operator commuting with the angle-dependent $\textrm{U}(1)$ charge. 
Following Ref.~\cite{Moult:2025njc}, we extend the in-in
Ward-identity argument from two to $N$ insertions. For connected correlators, this gives
\begin{equation}
\begin{split}
\Big\langle\prod_{i=1}^NQ^+_{S,i}\Big\rangle_c&=(-1)^N\Big\langle\prod_{i=1}^NQ^+_{H,i}\Big\rangle_c+\delta_{N,2}\,G_{12}\,,\\
G_{ij}&=\langle Q^-_{S,i}Q^-_{S,j}\rangle_{\rm vac}\,.
\label{eq:ward}
\end{split}
\end{equation}
Key steps of this derivation are discussed in \cite{SM}.


We have reduced memory correlators to computing charge correlators~\cite{Hofman:2008ar, Lee:2023npz, Monni:2025zyv}, see~\cite{Moult:2025njc, Herrmann:2026kue} for the analogous gravitational case.
Writing the charge correlators as weighted cross sections and inverting the divergence yields the moments of the memory reconstructed from the hard charges, denoted by $\langle \cdots\rangle_H$. For a theory of massless charged particles we have
\begin{equation}
\begin{split}
\Big\langle\prod^N_{i=1}\Delta A_{z_i}\Big\rangle_H=\frac{(-\alpha)^N}{\sigma_{\rm tot}}&\sum_X\int d\Phi_X\,|\mathcal{M}_{\mathcal{O}\to X}|^2\\
&\times\prod^{N}_{i=1} \Big[\sum_{a\in X}\frac{q_a}{z_i-z_a}\Big]\,,
\end{split}
\label{eq:moments}
\end{equation}
with $A_z(z_i)\equiv A_{z_i}$ and $d\Phi_X$ the Lorentz-invariant phase space measure including the momentum-conserving delta function and $\sigma_{\rm tot}$ the same integral without the weight factor. Memory insertions of opposite helicity are simply obtained by complex conjugating the weights. For $N\geq3$, the connected part coincides with the connected correlators of the full electric memory. At two points, the latter receives the additional contribution $G_{ij}$ from incoming soft vacuum fluctuations.

\emph{The electromagnetic current.}---We consider a lepton pair produced by the electromagnetic current and work in the rest frame of the source,
\begin{equation}
J^\mu = :\bar\psi\gamma^\mu\psi:~,\qquad q^\mu=(Q,\mathbf 0)\,.
\end{equation}

We average over the three spatial polarizations of the current such that the positive lepton is equally likely to emerge in any direction at tree level~\cite{SM}. 

For massless leptons, the memory of a tree level process is fixed once we know the direction $w$ of the positive charge as the negative charge is at the antipodal point $w_A= -1/\bar{w}$. In terms of the tree-level opening-angle distribution $\mathcal{D}(s_{ab})=\delta(1-s_{ab})$ of the charged pair, with $s_{ab}=\sin^2\frac{\theta_{ab}}{2}$, the memory correlator is given by
\begin{equation}
\begin{split}
    \Big\langle\prod^{N}_{i=1}\Delta A_{z_i}\Big\rangle_H=(-\alpha)^{N}&\int\frac{d\Omega_ad\Omega_b}{(4\pi)^2} \mathcal{D}(s_{ab})\label{eq:QCN} \\
\times \prod^N_{i=1}&\left(\frac{1}{z_i-z_a}-\frac{1}{z_i-z_b}\right)\,,
\end{split}
\end{equation}
which is just an average of the distribution times the memory weights. A derivation of this formula is provided in~\cite{SM}. For anti-holomorphic insertions, the respective memory kernels are conjugated.

Since reversing the pair axis flips the sign of the charges, isotropy implies that all odd moments vanish. For even $N$ and distinct, non-antipodal detector directions, partial fraction decomposition yields
\begin{equation}
\begin{split}
\Big\langle\prod_i\Delta A_{z_i}\Big\rangle_H&=\alpha^{N-2}\sum_{k<l}R_{kl}z^2_{kl}\langle\Delta A_{z_k}\Delta A_{z_l}\rangle_H\,,\\
R_{kl}&=\frac{(z_k-z_l)^{N-4}}{\prod_{m\neq k,l}(z_m-z_k)(z_l-z_m)}\,.
\label{eq:Npoint}
\end{split}
\end{equation}
Therefore, higher-point holomorphic correlators are fully fixed in terms of the two-point correlator~\cite{SM}, and transcendentality does not grow with $N$. 
The connected correlators then follow by subtracting disconnected contributions. At tree level, applying the antipodal map to the insertions in \eqref{eq:Npoint} produces mixed-helicity correlators.

At two points, the equal- and opposite-helicity correlators are
\begin{equation}
\begin{split}
\langle\Delta A_{z_1}\Delta A_{z_2}\rangle_H=2\alpha^2\frac{\log(1-s_{12})}{z_{12}^2}\,,\\
\langle\Delta A_{z_1}\Delta A_{\bz_2}\rangle_H=-2\alpha^2\frac{\log s_{12}}{(1+z_1\bz_2)^2}\,.
\end{split}
\label{eq:2pt}
\end{equation}
The connected two-point memory correlator contains the incoming vacuum sector $G_{ij}$. For opposite helicities, this vacuum is distributional, proportional to $\delta^{(2)}(z_{12})$, while for equal helicities, the electric projection adds a vacuum double pole $4\pi c\alpha^2/z_{12}^2$, whose coefficient $c\le0$ depends on the prescription of the soft regularization.

\begin{figure}[t]
\centering
\includegraphics[scale=0.4]{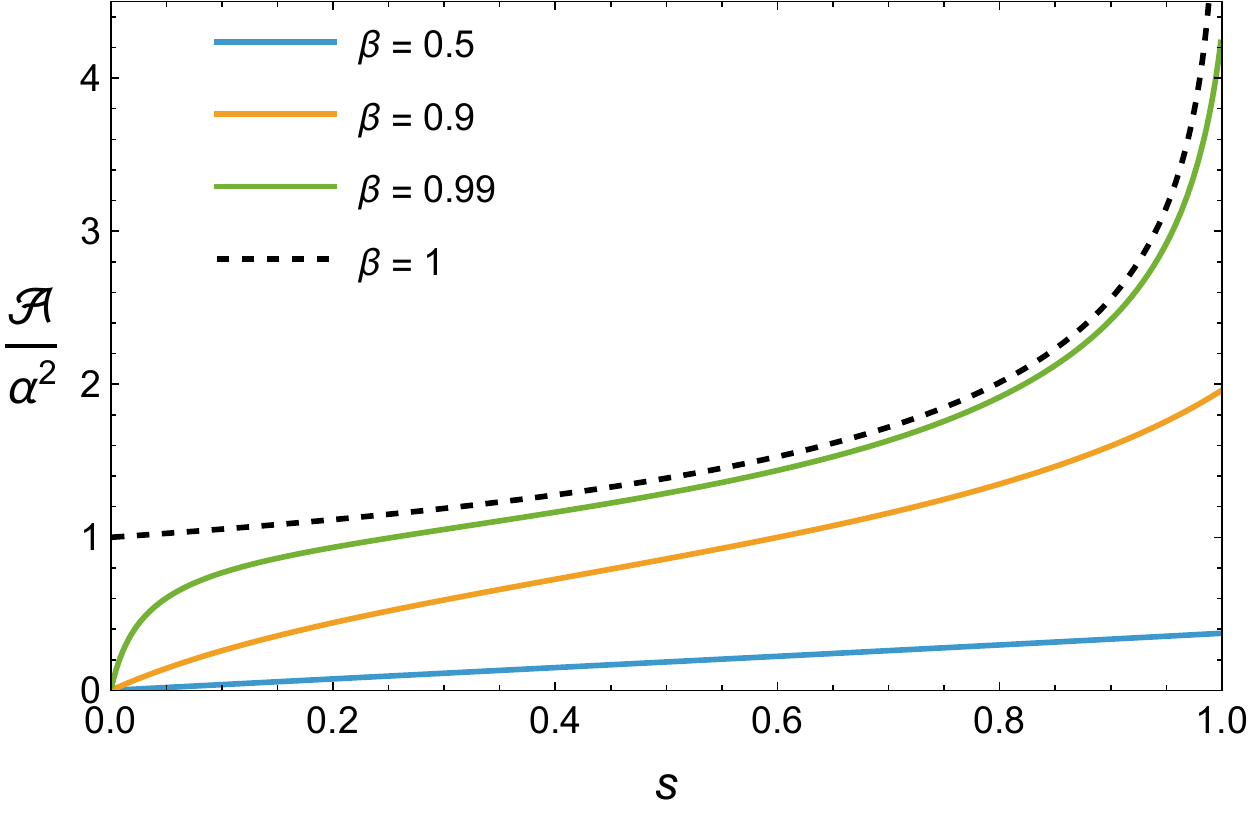}
\caption{The massive two-point function, normalized to $\alpha^2$. The dashed curve is the massless limit. Finite mass smooths the coincident region. Increasing the speed approaches the massless curve. The opposite-helicity function follows by reflecting the angle, as in Eq.~\eqref{eq:massive}.}
\label{fig:massive}
\end{figure}

\emph{Mass and soft radiation.}---A massive pair has opposite velocities $\pm\vec{v}$ in the source rest frame, with $\beta=|\vec{v}|=\sqrt{1-\frac{4m^2}{Q^2}}$.
The point-like charge profile is replaced by a smooth Coulomb profile, whose angular width decreases as the lepton is boosted~\cite{Campiglia:2015qka,Kapec:2015ena}, leading to $\Delta A_z=-\alpha q \partial_z \ln (1-\beta \hat{v}\cdot \hat{n})$ instead of~\eqref{eq:memory} and a corresponding change in the weights in \eqref{eq:moments}. One finds
\begin{align}
\langle\Delta A_{z_1}\Delta A_{z_2}\rangle_H={}&-\frac{2}{d_1d_2}\frac{\bz_{12}}{z_{12}}\,\mathcal{A}(s,\beta)\,,\nonumber\\
\langle\Delta A_{z_1}\Delta A_{\bz_2}\rangle_H={}&-\frac{2}{d_1d_2}\frac{1+\bz_1z_2}{1+z_1\bz_2}\,\mathcal{B}(s,\beta)\,,\nonumber\\
\mathcal{B}(s,\beta)={}&-\mathcal{A}(1-s,\beta)\,,
\label{eq:massive}
\end{align}
with $d_i=1+|z_i|^2$. The explicit elementary function is given in the Supplemental Material~\cite{SM}. It recovers the massless answer at fixed angular separation. Near antipodal or coincident points, however, finite mass resolves a region of order the inverse boost (Fig.~\ref{fig:massive}). The massive function is smooth there, while the massless limit leads to a singularity at these points.

Noteworthy, the coincident correlator has a direct physical meaning. The leading soft photon theorem relates the radiated photon number to the square of the memory through
\begin{equation}
\lim_{\omega \to 0} \frac{dN_\gamma}{d\log\omega\,d\Omega}=\frac{1}{\pi e^2}\,\Delta A_A\Delta A^A\,.
\label{eq:NtoS}
\end{equation}
This relation can be understood as deriving from a mixing between the photon number operator and the product of two memory operators. As shown in~\cite{Gonzalez:2025ene} (see also~\cite{Chang:2025zib}), the generalized energy operator measuring $E^{a}$ of outgoing photons develops a soft pole at $a=0$, with residue proportional to the right hand side of Equation~\eqref{eq:NtoS}. Subtracting the pole to define a renormalized operator leads then to the flow equation~\eqref{eq:NtoS}. 

For a massive pair, the coincident hard two-point function is finite and gives
\begin{equation}
\lim_{\omega \to 0}\Big\langle\frac{dN_\gamma}{d\log\omega}\Big\rangle_{H}=\frac{2\alpha}{\pi}\Big[\frac{1+\beta^2}{\beta}\arctanhh\beta-1\Big]\,.
\label{eq:softphotons}
\end{equation}
At high energy, boosting the pair increases the soft photon number logarithmically, with the lepton mass regulating the collinear region. The collinear enhancement of memory correlations scales as $\log(Q^2/m^2)$ and is thus the same enhancement that appears in soft photon radiation. Since the average memory vanishes for this source, the radiation is entirely encoded in its two-point function.

Equation~\eqref{eq:NtoS} provides an interesting interpretation for correlators of memory operators with pairwise coincident points as describing the change in angular correlation of soft photon countings as the soft scale is varied. While these coincident points are non-singular in the massive case, in the massless case one expects an additional necessary renormalization of the photon operator due to collinear divergences~\cite{Caron-Huot:2022eqs}. We will explore these issues in more detail in the companion paper~\cite{Diaz:2026emc}.

\emph{First radiative correction.}--- We calculate the first correction to the memory correlator for massless particles. The photon is neutral, so it enters the reconstructed memory only through the recoil of the charged pair, and equation~\eqref{eq:QCN} still holds. At NLO, the memory correlator is completely fixed by 
\begin{equation}
\cD(s_{ab})=\delta(1-s_{ab})+\frac{\alpha}{\pi}\big[\cP(s_{ab})\big]_+
+O(\alpha^2)\,,
\label{eq:Dist}
\end{equation}
where $\cP(s_{ab})$ comes from photons emitted by the two charged legs~\cite{SM}. When the photon is soft or collinear to one of the charges, the pair is back-to-back, $s_{ab}=1$. Normalizing to the total rate gives the plus prescription at this endpoint, so the correlators are finite for $0<s_{12}<1$ at this order. At the next relative order, however, wide-angle soft charged pairs can introduce infrared sensitivity~\cite{Monni:2025zyv}.

Similarly, equation~\eqref{eq:Npoint} continues to hold at NLO, for both purely holomorphic or anti-holomorphic insertions. 
 Recoil, however, breaks the antipodal map relation for mixed-correlators.    This means that the partial fraction decompositions leading to~\eqref{eq:Npoint} have to be done separately and higher-order functions may appear. To first radiative order, with $s=s_{12}$,
\begin{equation}
\begin{split}
\langle\Delta A_{z_1}\Delta A_{z_2}\rangle_H
&=\frac{2\alpha^2}{z_{12}^2}
{}\Bigg[\!\log(1\!-\!s)\!+\frac{\alpha}{\pi}\!
\int_0^s\!\!\frac{dy\,\,\cR(y)}{1-y}\Bigg],\\
\langle\Delta A_{z_1}\Delta A_{\bz_2}\rangle_H
&=-\frac{2\alpha^2}{(1+z_1\bz_2)^2} \label{eq:Oalpha}\\
&\qquad\times\Bigg[\log s+\frac{\alpha}{\pi}
\int_s^1\frac{dy}{y}\,\cR(y)\Bigg]\,,
\end{split}
\end{equation}
where $\mathcal{R}(s)=\int^{s}_0\, dy\, \mathcal{P}(y).$
Near coincidence $s=0$, $\cR(s)=s/3+O(s^2)$, so the leading correction in the equal-helicity correlator is $\tfrac{\alpha s^2}{6 \pi}$. Similarly, in the opposite-helicity case, the correction tends to $\tfrac{\alpha}{\pi}\big(2\zeta(3)-\tfrac{11}{8}\big)$, with transcendentality three, leaving the coefficient of $\log s$ unchanged. 

Near the antipode, we probe the back-to-back region where recoil produces a double Sudakov logarithm in $\mathcal{R}(s)$. Writing $L=-\log(1-s)$, we find
\begin{equation}
\begin{split}
&\cR(s)=\tfrac12L^2-\tfrac32L+\tfrac74+\tfrac{\pi^2}{6}
+O\big((1-s)L\big)\,,\\
&\frac{z_{12}^2}{2\alpha^2}
\langle\Delta A_{z_1}\Delta A_{z_2}\rangle_H
=-L+\frac{\alpha}{\pi}
\Big[\tfrac16L^3-\tfrac34L^2+O(L)\Big]\,.
\end{split}
\label{eq:Sudakov}
\end{equation}
Therefore, at generic angles, the leading result contains only single logarithms, while the first correction produces trilogarithms. 
Interestingly, when $(\alpha/\pi)L^2 \sim 1$, the cubic logarithm is as large as the LO term. This means that multiple soft emissions contribute with terms of the same size, $(\alpha/\pi)^nL^{2n}$, which must be resummed to all orders. The fact that the back-to-back limit probes soft physics has been studied extensively~\cite{Korchemsky:2019nzm, Gao:2019ojf, Herrmann:2024yai, Chicherin:2025keq}.

\emph{Celestial free boson.}---The observables obtained here are reproduced by a free boson $\phi$ on the celestial sphere,
\begin{equation}
\langle\phi(z)\phi(w)\rangle=-k\log s(z,w)\,,
\label{eq:boson}
\end{equation}
with current $J=\partial_z\phi$ and vertex operators $V_q=\,:\!e^{-q\phi/k}\!:$. Their operator products,
\begin{equation}
J(z)J(w)\sim-\frac{k}{(z-w)^2}\,,\quad J(z)V_q(w)\sim\frac{q}{z-w}\,V_q(w)\,,
\label{eq:OPE}
\end{equation}
encode the soft photon theorem on the sphere~\cite{Nande:2017dba,Himwich:2020rro}.
Vertices with opposite charge $q=\pm1$ at arbitrary points $z_{a},z_{b}$ reproduce the memory kernel of an oppositely charged pair $\langle V_1(z_a)J(z)V_{-1}(z_b)\rangle/\langle V_1V_{-1}\rangle=1/(z-z_a)-1/(z-z_b)$. We can then define a generating functional for the fixed vertices 
\begin{equation}
    Z_{ab}[\,\vec{t}\,\,]=\frac{\langle V_1(z_a) \prod_i : e^{-\alpha t_i J(z_i)}: V_{-1}(z_b)\rangle }{\langle V_1(z_a) V_{-1}(z_b)\rangle},
\end{equation}
such that standard Wick contraction leads to
\begin{align}
Z_{ab}[\,\vec{t}\,\,]=e^{\alpha^2\sum_{i<j}t_it_jK_{ij}}\,e^{-\alpha\sum_it_i\left[ (z_i-z_a)^{-1}-(z_i-z_b)^{-1}\right]}\,,
\end{align}
with $
K_{ij}=-k/z^2_{ij}$. As in the previous Section, the distribution of the vertices on the celestial sphere is fixed by \eqref{eq:Dist}. Averaging the position of the vertices against this distribution leads to 
\begin{align}
    Z[\,\vec{t}\,\,]&\!=\int^{1}_0\,du\, \mathcal{D}(u)\int\frac{d \Omega_a}{4\pi}\int^{2\pi}_0 \frac{d\varphi_{ab}}{2\pi}Z_{ab}[\,\vec{t}\,\,]|_{s_{ab}=u},
    \label{eq:protagonist}
\end{align}
where $\varphi_{ab}$ is the azimuthal angle of $b$ around the direction of $a$. Derivatives at $\vec{t}=0$ reproduce Eqs.~\eqref{eq:QCN}, and~\eqref{eq:Npoint} up to the inclusion of the vacuum contribution due to $K_{ij}$, which reproduces the incoming vacuum double pole with $k=-4\pi c$. We have restricted ourselves to the case of purely holomorphic insertions. However, mixed correlators of holomorphic and anti-holomorphic memories can be organized using the same ingredients.

For the massive case at tree-level, we can give a similar representation by replacing the point-like vertices by extended ones,
\begin{equation}
V_{\beta,a}=\,:\!\exp\Big[-\frac1k\int d^2\nh\,\rho_{\beta,a}(\nh)\,\phi(\nh)\Big]\!:\,,
\label{eq:extended}
\end{equation}
where $\rho_{\beta,a}$ is the difference of the massive profiles $1/[4\pi(-Y_\pm\cdot\ell)^2]$ of the two charges, with $Y_\pm^\mu=\gamma(1,\pm\beta\nh_a)$ and $\ell^\mu=(1,\nh)$. 

The celestial CFT representation provides a compact form to organize the angular structure of memory correlators. Beyond tree level, one would expect a more complicated pattern, since radiative photon emission changes both direction and the velocities of the charged particles.
Notably, the two-dimensional description is preserved at NLO using the generating functional \eqref{eq:protagonist} with $\mathcal{D}(u)$ defined by \eqref{eq:Dist} for the massless case. In fact, even in the massive case, it is possible to write down an analogous generating function at NLO by integrating over the mass-shell~\cite{Diaz:2026emc}.
It should be emphasized that the vertex distribution is an external input from the underlying scattering process. It would be interesting to see whether one can derive constraints on this additional input from the celestial CFT perspective.

\emph{Outlook.}---The memory correlators obtained here are predictions of the ${\rm U}(1)$ Ward identity in electrodynamics. We reduced their computation to correlators of charge detectors. While being potentially measurable independently, we also proposed an independent test of pair-wise coincident memory correlators as counting soft photons at different angles as the soft scale is varied. The Ward identity therefore connects correlators of memory operators, charge detectors, and energy-dependence of soft photon countings. For the latter, one can imagine a setup where the memory correlations are obtained by counting photons within a fixed small-energy window in two narrow cones around directions $\hat n_1$ and $\hat n_2$. 
We will present this perspective in more detail in the companion paper~\cite{Diaz:2026emc}, together with higher-order pair emission, a complete celestial description, and the extension to color memory correlations. 

\emph{Note added.}---While this work was being finalized, gravitational memory correlators for scalar decay into gravitons were computed at leading order in \cite{Herrmann:2026kue}.
\vspace{0.5cm}
\begin{acknowledgments}
\emph{Acknowledgments.}
The work of HG and OL is funded by the ANID FONDECYT grant 1230853. The work of JS was supported by a Postdoctoral Research Fellowship granted by the F.R.S.-FNRS (Belgium) and is currently supported by the Austrian Science Fund (FWF) [Grants DOI : \href{https://www.fwf.ac.at/en/research-radar/10.55776/P36619}{10.55776/P36619}, \href{https://www.fwf.ac.at/en/research-radar/10.55776/PAT1871425}{10.55776/PAT1871425}]. The work of FD is supported by ANID FONDECYT grant 3260880.
\end{acknowledgments}
\bibliography{memory}

@misc{Diaz:2026emc,
    author = "Diaz, Felipe and Gonz{\'a}lez, Hern{\'a}n A. and Labrin, Oriana and Salzer, Jakob",
    title = "{Electromagnetic memory correlators: Ward identities, massive charges and radiative corrections}",
    note = "In preparation",
    year = "2026"
}

@misc{SM,
    note = "See Supplemental Material for the Dirac source normalization, the connected Ward identity, the multipoint proof, the exact massive two-point function, and the fermionic photon-emission calculation."
}

@article{Low:1958sn,
    author = "Low, F. E.",
    title = "{Bremsstrahlung of very low-energy quanta in elementary particle collisions}",
    doi = "10.1103/PhysRev.110.974",
    journal = "Phys. Rev.",
    volume = "110",
    pages = "974--977",
    year = "1958"
}

@article{Korchemsky:2019nzm,
    author = "Korchemsky, G. P.",
    title = "{Energy correlations in the end-point region}",
    eprint = "1905.01444",
    archivePrefix = "arXiv",
    primaryClass = "hep-th",
    reportNumber = "IPhT-T19/041",
    doi = "10.1007/JHEP01(2020)008",
    journal = "JHEP",
    volume = "01",
    pages = "008",
    year = "2020"
}

@article{Gao:2019ojf,
    author = "Gao, AnJie and Li, Hai Tao and Moult, Ian and Zhu, Hua Xing",
    title = "{Precision QCD Event Shapes at Hadron Colliders: The Transverse Energy-Energy Correlator in the Back-to-Back Limit}",
    eprint = "1901.04497",
    archivePrefix = "arXiv",
    primaryClass = "hep-ph",
    doi = "10.1103/PhysRevLett.123.062001",
    journal = "Phys. Rev. Lett.",
    volume = "123",
    number = "6",
    pages = "062001",
    year = "2019"
}

@misc{Lee:2023npz,
    author = "Lee, Kyle and Moult, Ian",
    title = "{Energy Correlators Taking Charge}",
    eprint = "2308.00746",
    archivePrefix = "arXiv",
    primaryClass = "hep-ph",
    reportNumber = "MIT-CTP-5590",
    month = "8",
    year = "2023"
}

@misc{Herrmann:2026kue,
    author = "Herrmann, Enrico and Moult, Ian and Oertel, Brett and Parra-Martinez, Julio and Sahoo, Biswajit",
    title = "{Gravitational Memory Correlators and BMS Symmetry}",
    eprint = "2611.18827",
    archivePrefix = "arXiv",
    primaryClass = "hep-th",
    month = "11",
    year = "2026"
}

@article{Caron-Huot:2023vxl,
    author = "Caron-Huot, Simon and Giroux, Mathieu and Hannesdottir, Holmfridur S. and Mizera, Sebastian",
    title = "{What can be measured asymptotically?}",
    eprint = "2308.02125",
    archivePrefix = "arXiv",
    primaryClass = "hep-th",
    doi = "10.1007/JHEP01(2024)139",
    journal = "JHEP",
    volume = "01",
    pages = "139",
    year = "2024"
}

@article{Chicherin:2025keq,
    author = "Chicherin, Dmitry and Korchemsky, Gregory P. and Sokatchev, Emery and Zhiboedov, Alexander",
    title = "{Energy correlators in four-dimensional gravity}",
    eprint = "2512.23791",
    archivePrefix = "arXiv",
    primaryClass = "hep-th",
    reportNumber = "CERN-TH-2025-266, LAPTH-060/25",
    doi = "10.1007/JHEP07(2026)275",
    journal = "JHEP",
    volume = "07",
    pages = "275",
    year = "2026"
}

@article{Herrmann:2024yai,
    author = "Herrmann, Enrico and Kologlu, Murat and Moult, Ian",
    title = "{Energy correlators in perturbative quantum gravity}",
    eprint = "2412.05384",
    archivePrefix = "arXiv",
    primaryClass = "hep-th",
    doi = "10.1007/JHEP07(2026)126",
    journal = "JHEP",
    volume = "07",
    pages = "126",
    year = "2026"
}

@article{Chang:2025zib,
    author = "Chang, Cyuan-Han and Chen, Hao and Simmons-Duffin, David and Zhu, Hua Xing",
    title = "{Seeing through the confinement screen: DGLAP/BFKL mixing and light-ray matching in QCD}",
    eprint = "2506.06431",
    archivePrefix = "arXiv",
    primaryClass = "hep-th",
    reportNumber = "MIT-CTP 5876, CALT-TH 2025-018",
    doi = "10.1007/JHEP02(2026)251",
    journal = "JHEP",
    volume = "02",
    pages = "251",
    year = "2026"
}

@article{Hu:2023geb,
    author = "Hu, Yangrui and Pasterski, Sabrina",
    title = "{Detector operators for celestial symmetries}",
    eprint = "2307.16801",
    archivePrefix = "arXiv",
    primaryClass = "hep-th",
    doi = "10.1007/JHEP12(2023)035",
    journal = "JHEP",
    volume = "12",
    pages = "035",
    year = "2023"
}

@article{Caron-Huot:2022eqs,
    author = "Caron-Huot, Simon and Kologlu, Murat and Kravchuk, Petr and Meltzer, David and Simmons-Duffin, David",
    title = "{Detectors in weakly-coupled field theories}",
    eprint = "2209.00008",
    archivePrefix = "arXiv",
    primaryClass = "hep-th",
    reportNumber = "CALT-TH 2022-31",
    doi = "10.1007/JHEP04(2023)014",
    journal = "JHEP",
    volume = "04",
    pages = "014",
    year = "2023"
}

@article{Campiglia:2015qka,
    author = "Campiglia, Miguel and Laddha, Alok",
    title = "{Asymptotic symmetries of QED and Weinberg{\textquoteright}s soft photon theorem}",
    eprint = "1505.05346",
    archivePrefix = "arXiv",
    primaryClass = "hep-th",
    doi = "10.1007/JHEP07(2015)115",
    journal = "JHEP",
    volume = "07",
    pages = "115",
    year = "2015"
}

@misc{Oertel:2026oqv,
    author = "Oertel, Brett",
    title = "{Finite-time memory detectors and fully constraining Faddeev-Kulish dressings in QED and gravity}",
    eprint = "2605.06774",
    archivePrefix = "arXiv",
    primaryClass = "hep-th",
    month = "5",
    year = "2026"
}

@article{Bieri:2013hqa,
    author = "Bieri, Lydia and Garfinkle, David",
    title = "{An electromagnetic analogue of gravitational wave memory}",
    eprint = "1307.5098",
    archivePrefix = "arXiv",
    primaryClass = "gr-qc",
    doi = "10.1088/0264-9381/30/19/195009",
    journal = "Class. Quant. Grav.",
    volume = "30",
    pages = "195009",
    year = "2013"
}

@article{Donnay:2018neh,
    author = "Donnay, Laura and Puhm, Andrea and Strominger, Andrew",
    title = "{Conformally Soft Photons and Gravitons}",
    eprint = "1810.05219",
    archivePrefix = "arXiv",
    primaryClass = "hep-th",
    doi = "10.1007/JHEP01(2019)184",
    journal = "JHEP",
    volume = "01",
    pages = "184",
    year = "2019"
}

@article{Fuentealba:2025ekj,
    author = "Fuentealba, Oscar and Henneaux, Marc",
    title = "{Logarithmic angle-dependent gauge transformations at null infinity}",
    eprint = "2504.05385",
    archivePrefix = "arXiv",
    primaryClass = "hep-th",
    doi = "10.1007/JHEP07(2025)112",
    journal = "JHEP",
    volume = "07",
    pages = "112",
    year = "2025"
}

@article{He:2014cra,
    author = "He, Temple and Mitra, Prahar and Porfyriadis, Achilleas P. and Strominger, Andrew",
    title = "{New Symmetries of Massless QED}",
    eprint = "1407.3789",
    archivePrefix = "arXiv",
    primaryClass = "hep-th",
    doi = "10.1007/JHEP10(2014)112",
    journal = "JHEP",
    volume = "10",
    pages = "112",
    year = "2014"
}

@misc{Oertel:2026wsm,
    author = "Oertel, Brett and Moult, Ian and Pasterski, Sabrina",
    title = "{Asymptotic charges as detectors and the memory effect in massive QED and perturbative quantum gravity}",
    eprint = "2604.19866",
    archivePrefix = "arXiv",
    primaryClass = "hep-th",
    month = "4",
    year = "2026"
}

@book{Ashtekar:1987tt,
    author = "Ashtekar, A.",
    title = "{Asymptotic Quantization: Based on 1984 Naples Lectures}",
    year = "1987"
}

@article{Gonzalez:2025ene,
    author = "Gonz{\'a}lez, Hern{\'a}n A. and Salzer, Jakob",
    title = "{Energy detectors and asymptotic symmetries}",
    eprint = "2510.27348",
    archivePrefix = "arXiv",
    primaryClass = "hep-th",
    doi = "10.1007/JHEP07(2026)241",
    journal = "JHEP",
    volume = "07",
    pages = "241",
    year = "2026"
}

@article{Himwich:2020rro,
    author = "Himwich, Elizabeth and Narayanan, Sruthi A. and Pate, Monica and Paul, Nisarga and Strominger, Andrew",
    title = "{The Soft $\mathcal{S}$-Matrix in Gravity}",
    eprint = "2005.13433",
    archivePrefix = "arXiv",
    primaryClass = "hep-th",
    doi = "10.1007/JHEP09(2020)129",
    journal = "JHEP",
    volume = "09",
    pages = "129",
    year = "2020"
}

@article{Kapec:2015ena,
    author = "Kapec, Daniel and Pate, Monica and Strominger, Andrew",
    title = "{New Symmetries of QED}",
    eprint = "1506.02906",
    archivePrefix = "arXiv",
    primaryClass = "hep-th",
    doi = "10.4310/ATMP.2017.v21.n7.a7",
    journal = "Adv. Theor. Math. Phys.",
    volume = "21",
    pages = "1769--1785",
    year = "2017"
}

@article{Nande:2017dba,
    author = "Nande, Anjalika and Pate, Monica and Strominger, Andrew",
    title = "{Soft Factorization in QED from 2D Kac-Moody Symmetry}",
    eprint = "1705.00608",
    archivePrefix = "arXiv",
    primaryClass = "hep-th",
    doi = "10.1007/JHEP02(2018)079",
    journal = "JHEP",
    volume = "02",
    pages = "079",
    year = "2018"
}

@article{Pasterski:2015zua,
    author = "Pasterski, Sabrina",
    title = "{Asymptotic Symmetries and Electromagnetic Memory}",
    eprint = "1505.00716",
    archivePrefix = "arXiv",
    primaryClass = "hep-th",
    doi = "10.1007/JHEP09(2017)154",
    journal = "JHEP",
    volume = "09",
    pages = "154",
    year = "2017"
}

@article{Weinberg:1965nx,
    author = "Weinberg, Steven",
    title = "{Infrared photons and gravitons}",
    doi = "10.1103/PhysRev.140.B516",
    journal = "Phys. Rev.",
    volume = "140",
    pages = "B516--B524",
    year = "1965"
}

@misc{Strominger:2017zoo,
      author         = "Strominger, Andrew",
      title          = {Lectures on the Infrared Structure of Gravity and Gauge Theory},
      year           = "2017",
      eprint         = "1703.05448",
      archivePrefix  = "arXiv",
      primaryClass   = "hep-th",
      SLACcitation   = "%%CITATION = ARXIV:1703.05448;%%"
}

@article{Kravchuk:2018htv,
    author = "Kravchuk, Petr and Simmons-Duffin, David",
    title = "{Light-ray operators in conformal field theory}",
    eprint = "1805.00098",
    archivePrefix = "arXiv",
    primaryClass = "hep-th",
    reportNumber = "CALT-TH 2018-018",
    doi = "10.1007/JHEP11(2018)102",
    journal = "JHEP",
    volume = "11",
    pages = "102",
    year = "2018"
}

@article{Basham:1978zq,
    author = "Basham, C. L. and Brown, L. S. and Ellis, S. D. and Love, S. T.",
    title = "{Energy Correlations in electron-Positron Annihilation in Quantum Chromodynamics: Asymptotically Free Perturbation Theory}",
    reportNumber = "RLO-1388-761",
    doi = "10.1103/PhysRevD.19.2018",
    journal = "Phys. Rev. D",
    volume = "19",
    pages = "2018",
    year = "1979"
}

@article{Hofman:2008ar,
    author = "Hofman, Diego M. and Maldacena, Juan",
    title = "{Conformal collider physics: Energy and charge correlations}",
    eprint = "0803.1467",
    archivePrefix = "arXiv",
    primaryClass = "hep-th",
    doi = "10.1088/1126-6708/2008/05/012",
    journal = "JHEP",
    volume = "05",
    pages = "012",
    year = "2008"
}

@misc{Moult:2025nhu,
    author = "Moult, Ian and Zhu, Hua Xing",
    title = "{Energy Correlators: A Journey From Theory to Experiment}",
    eprint = "2506.09119",
    archivePrefix = "arXiv",
    primaryClass = "hep-ph",
    month = "6",
    year = "2025"
}

@article{Monni:2025zyv,
    author = "Monni, Pier Francesco and Vita, Gherardo and Xu, Zhen and Zhu, Hua Xing",
    title = "{Edge of Safety: Charge-Charge Correlation in the Back-to-Back Limit}",
    eprint = "2508.00977",
    archivePrefix = "arXiv",
    primaryClass = "hep-ph",
    reportNumber = "CERN-TH-2025-150",
    doi = "10.1103/3pqk-qr64",
    journal = "Phys. Rev. Lett.",
    volume = "137",
    number = "3",
    pages = "031901",
    year = "2026"
}

@article{Moult:2025njc,
    author = "Moult, Ian and Narayanan, Sruthi A. and Pasterski, Sabrina",
    title = "{Memory correlators and Ward identities in the 'in-in' formalism}",
    eprint = "2512.02825",
    archivePrefix = "arXiv",
    primaryClass = "hep-th",
    doi = "10.1007/JHEP09(2026)239",
    journal = "JHEP",
    volume = "09",
    pages = "239",
    year = "2026"
}

@article{Bieri:2023eme,
    author = "Bieri, Lydia and Garfinkle, David",
    title = "{An experiment to measure electromagnetic memory}",
    eprint = "2401.00234",
    archivePrefix = "arXiv",
    primaryClass = "gr-qc",
    journal = "Class. Quant. Grav.",
    volume = "41",
    pages = "225009",
    year = "2024",
    doi = "10.1088/1361-6382/ad8542"
}

@article{Zosso:2025eme,
    author = "Zosso, Jann",
    title = "{Enhancement of Electromagnetic Memory Effects}",
    eprint = "2507.09555",
    archivePrefix = "arXiv",
    primaryClass = "gr-qc",
    doi = "10.1103/hjrp-5cps",
    journal = "Phys. Rev. Lett.",
    volume = "135",
    number = "20",
    pages = "201602",
    year = "2025"
}

\clearpage
\onecolumngrid
\begin{center}
{\large\bfseries Supplemental Material}\\[4pt]
{\bfseries Electromagnetic Memory as a Collider Observable}
\end{center}
\setcounter{equation}{0}
\setcounter{figure}{0}
\renewcommand{\theequation}{S\arabic{equation}}
\renewcommand{\theHequation}{S\arabic{equation}}
\renewcommand{\thefigure}{S\arabic{figure}}

\noindent Further details are given in the companion paper~\cite{Diaz:2026emc}.

\section{The connected Ward identity}
To compute the soft charge correlators, we apply the in-in
Ward identity to each of the $N$ outgoing soft insertions.  Since the charge commutes with the local source and the incoming vacuum contains no hard charges, we have $Q_H^-|0\rangle=0$, therefore each $Q_S^+$ acts as $Q_S^- - Q_H^+$. Expanding the product will produce a sum of  correlators of the form
\begin{equation}
\big\langle O^+ Q^-_{S,1}\cdots Q^-_{S,m}\big\rangle\,,
\qquad
O^+|X\rangle = F(X)|X\rangle\,,
\label{OplusQs}
\end{equation}
with $m\leq N$ insertions of the soft incoming charges, $O^+$ denoting the product of charges 
defined at the future, and $F(X)$ their eigenvalue on a given state $X$.

To evaluate these mixed correlators, we adapt the inclusive cancellation argument given in~\cite{Moult:2025njc} to soft photons. When the $m$ soft charges act on the vacuum, they organize by contracting in pairs. Each uncontracted insertion creates a soft photon $\ell$ that can either be absorbed in the amplitude or pass the cut without interacting. In the latter case, $\ell$ is emitted in the conjugate amplitude. Soft photon theorem~\cite{Weinberg:1965nx} produces these two contributions with the same soft factors but opposite signs. At fixed hard momenta, their detector weights are equal, because the photon is neutral. The two contributions therefore cancel in the sum over outgoing states whenever at least one insertion is uncontracted. Only complete vacuum pairings survive, giving
\begin{equation}
\left\langle O^+\prod_{i=1}^{m}Q^-_{S,i}\right\rangle
=
\langle O^+\rangle
\begin{cases}
\displaystyle
\sum_{\mathcal P}
\prod_{(i,j)\in\mathcal P}G_{ij},
& m\ \text{even},\\[6pt]
0,
& m\ \text{odd},
\end{cases}
\label{eq:app-mixed-factorization}
\end{equation}
where $\mathcal P$ is the set of complete pairings and $G_{ij}=\langle Q^-_{S,i}Q^-_{S,j}\rangle_{\rm vac}\,$. In the connected $N$-point correlator, a vacuum pair vanishes from all other insertions, but it contributes only for $N=2$, as $G_{12}$. Thus, the purely charge correlator ($m=0$) contributes to the connected outgoing charge correlator with the factor $(-1)^N$. Together, these give Eq.~\eqref{eq:ward} of the Letter.

\section{Massless multipoint formula}
We prove Equation~\eqref{eq:Npoint} for even $N\ge2$. 
Setting $z_{ij}=z_i-z_j$, the product of memory kernels can be written as
\begin{equation}
    \prod^{N}_{i=1}\left(\frac{1}{z_{ia}}-\frac{1}{z_{ib}}\right)=\frac{z^N_{ab}}{\prod_i z_{ia}z_{ib}}. 
\end{equation}
A partial-fraction decomposition yields
\begin{equation}
    \frac{z^{N-2}_{ab}}{\prod_i z_{ia}z_{ib}}=-\sum_{k,l}\frac{R_{kl}}{z_{ka}z_{lb}}\qquad R_{kl}=-\frac{z^{N-2}_{kl}}{\prod_{m\neq k}z_{mk}\prod_{n\neq l}z_{nl}}=\frac{z^{N-4}_{kl}}{\prod_{m\neq l,k}z_{mk}z_{ml}}\qquad R_{kl}=R_{lk}.
\end{equation}
Since the function vanishes as $O(z^{-2}_a)$ and $O(z^{-2}_b)$, the coefficients obey the constraints $\sum_{k} R_{kl}=\sum_{l}R_{kl}=0$.
Rearranging the sum using these identities, we have
\begin{equation}
    \frac{z^{N-2}_{ab}}{\prod_i z_{ia}z_{ib}}=-\sum_{k,l}\frac{R_{kl}}{z_{ka}z_{lb}}=\sum_{k<l} R_{kl} \left(\frac{1}{z_{ka}}-\frac{1}{z_{la}}\right)\left(\frac{1}{z_{kb}}-\frac{1}{z_{lb}}\right),
\end{equation}
and 
\begin{equation}
    \prod^{N}_{i=1}\left(\frac{1}{z_{ia}}-\frac{1}{z_{ib}}\right)=\sum_{k<l} R_{kl} z^2_{kl}\left(\frac{1}{z_{ka}}-\frac{1}{z_{kb}}\right)\left(\frac{1}{z_{la}}-\frac{1}{z_{lb}}\right).
\end{equation}
Inserting this identity into Equation~\eqref{eq:QCN} and recognizing the last two brackets as the product of memory kernels entering the two-point function, we obtain Equation~\eqref{eq:Npoint}. 
\section{Massless and Massive two-point function}
\subsection{Massless two-point function}
To compute the two-point function we need to evaluate
\begin{equation}
\langle \Delta A_z(z) \Delta A_z(0)\rangle_H=2\alpha^2\int\frac{d\Omega_ad\Omega_b}{(4\pi)^2} \mathcal{D}(s_{ab}) \frac{1}{z-z_b}\left(\frac{1}{z_a}-\frac{1}{z_b}\right) 
\end{equation}
with $\mathcal{D}(s_{ab})$ depending on 
\begin{equation}
    s_{ab}=\sin^2\frac{\theta_{ab}}{2}=\frac{|z_a-z_b|^2}{(1+|z_a|^2)(1+|z_b|^2)}
\end{equation}
where we used the symmetry in $a\leftrightarrow b$. We first center the $z_a$ integral around $z_b$ by
\begin{equation}
    z_a=\frac{z_b+\rho e^{i\varphi}}{1-\bar{z}_b \rho e^{i\varphi}}
\end{equation}
and set $y=\rho^2/(1+\rho^2)$. Then $s_{ab}=y$ and 
\begin{equation}
    \frac{d \Omega_a}{4\pi}=\frac{d^2z_a}{\pi(1+|z_a|^2)^2}=dy \frac{d\varphi}{2\pi}.
\end{equation}
The correlator simplifies to
\begin{equation}
  \langle \Delta A_z(z) \Delta A_z(0)\rangle_H=2\alpha^2\int\frac{d^2z_b}{\pi(1+|z_b|^2)}\frac{1}{z-z_b}\int^{1}_0 d y\,\mathcal{D}(y) \int^{2\pi}_0\frac{d \varphi}{2\pi} \left(\frac{1}{z_b+\rho e^{i\varphi}}-\frac{1}{z_b}\right).   
\end{equation}
Using
\begin{equation}
    \int^{2\pi}_0 \frac{d\varphi }{2\pi}\frac{1}{z_b-\rho e^{i\varphi}}=\frac{1}{z_b}\Theta(|z_b|-\rho).
\end{equation}
we find
\begin{equation}
  \langle \Delta A_z(z) \Delta A_z(0)\rangle_H=-2\alpha^2\int\frac{d^2z_b}{\pi(1+|z_b|^2)}\frac{1}{z_b(z-z_b)}\int^{1}_\frac{|z_b|^2}{(1+|z_b|^2)} d y\,\mathcal{D}(y).
\end{equation}
Evaluating again the contour integration one has for generic locations
\begin{equation}
    \langle \Delta A_z(z_1) \Delta A_z(z_2)\rangle_H=-\frac{2\alpha^2}{z^2_{12}}\int^{s_{12}}_0 \frac{d v}{1-v}\int^1_v dy \, \mathcal{D}(y). \label{eq:final2pt}
\end{equation}
The mixed two-point function follows a similar pattern, yielding
\begin{equation}
\langle\Delta A_z(z_1)\Delta A_{\bz}(z_2)\rangle_H
=\frac{2\alpha^2}{(1+z_1\bz_2)^2}
\int_{s_{12}}^1\frac{dv}{v}
\int_v^1dy\,\mathcal D(y).
\end{equation}
Inserting the Born value $\mathcal{D}(y) = \delta(1-y)$, gives the the tree-level result~\eqref{eq:2pt}.
\subsection{Massive two-point function}
\label{supp:massive}
For opposite velocities along the pair axis, the memory kernel and a Feynman-parameter master integral are
\begin{gather}
h_\beta(z,\nh_a)=\partial_z\log\frac{1-\beta\,\nh(z)\cdot\nh_a}{1+\beta\,\nh(z)\cdot\nh_a}\,,\\
\int\frac{d\Omega_a}{4\pi}\frac{1}{(1-\beta\nh_1\cdot\nh_a)(1-\beta\nh_2\cdot\nh_a)}=\int_0^1\frac{dx}{1-\beta^2+4\beta^2s\,x(1-x)}=\frac{1}{2} \mathscr{L}(s,\beta)\,.
\end{gather}
Partial fractions of the two kernels reduce their angular average to this integral and its crossed counterpart. Resolving the tensor numerators gives Eq.~\eqref{eq:massive}, with
\begin{equation}
\frac{\mathcal{A}(s,\beta)}{\alpha^2}=\frac{2\arctanhh\beta}{s\beta}-\frac{1-\beta^2}{2}\mathscr{L}(s,\beta)-\frac{2-s(1+\beta^2)}{2s}\mathscr{L}(1-s,\beta)\,,\qquad
\mathscr{L}(s,\beta)=\frac{2\arctanhh\sqrt{\beta^2s/[1-\beta^2(1-s)]}}{\beta\sqrt{s[1-\beta^2(1-s)]}}\,.
\end{equation}
The endpoints needed for the coincident soft-photon number are
\begin{equation}
\mathcal{A}(0,\beta)=0\,,\qquad \frac{\mathcal{A}(1,\beta)}{\alpha^2}=\frac{1+\beta^2}{\beta}\arctanhh\beta-1\,,\qquad \mathscr{L}(0,\beta)=\frac{Q^2}{2m^2}\,,\qquad \mathscr{L}(1,\beta)=\frac{2\arctanhh\beta}{\beta}\,.
\end{equation}
Contracting the coincident tensor in Eq.~\eqref{eq:massive}, integrating over the sphere and using the soft-photon phase space gives Eq.~\eqref{eq:softphotons}.

\section{Fermionic real emission and finite radiative terms}
We want to derive \eqref{eq:QCN}, holding to NLO, from the weighted cross-section expression \eqref{eq:moments}.
To NLO order, there is a single pair of charged leptons, fixing the form of the memory weights in \eqref{eq:QCN}. We can write
\begin{equation}
\begin{split}
\Big\langle\prod^N_{i=1}\Delta A_z(z_i)\Big\rangle_H=\frac{(-\alpha)^N}{\sigma_{\rm tot}}&\int d\Omega_a d\Omega_b\frac{d \sigma}{d \Omega_a d \Omega_b}\,\prod^{N}_{i=1} \Big[\frac{1}{z_i-z_a}-\frac{1}{z_i-z_b}\Big]\,,
\end{split}
\end{equation}
with 
\begin{equation}
    \frac{d \sigma}{d \Omega_a d \Omega_b}=\sum_X\int d\Phi_X\,|\mathcal{M}_{\mathcal{O}\to X}|^2\delta^{(2)}(\vec{n}_a,\vec{n}_+)\delta^{(2)}(\vec{n}_b,\vec{n}_-)
    \label{eq:sigma}
\end{equation}
where $\vec{n}_{\pm}$ are the directions of the charged particles in the matrix element. Due to rotational invariance of the source we further set
\begin{equation}
    \frac{1}{\sigma_{tot}}\frac{d \sigma}{d \Omega_a d \Omega_b}=\frac{\mathcal{D}(y)}{(4\pi)^2},\qquad y=\frac{1-\vec{n}_a\cdot\vec{n}_b}{2}=\sin^2\frac{\theta_{ab}}{2}, \qquad \int^{1}_0 d y\,\mathcal{D}(y)=1,
\end{equation}
which leads to equation~\eqref{eq:QCN}.

We compute $\mathcal{D}(y)$ to NLO. 
The squared two-point amplitude averaged over spatial polarizations of the source and the normalization yield
\begin{equation}
\overline{|\mathcal M_{J\to 2}|^2}=\frac43Q^2\,,\qquad \sigma_0=\frac{Q^2}{6\pi}\,,\qquad \frac{\overline{|\mathcal M_{J\to 2}|^2}d\Phi_2}{\sigma_0}=\frac{d\Omega_+}{4\pi}\,\qquad \hat{n}_+=-\hat{n}_-.
\end{equation}
Together with the identity $\delta^{(2)}(\hat{n}_a,-\hat{n}_b)=(4\pi)^{-1}\delta(1-y)$, one finds at leading order $\mathcal{D}(y)=\delta(1-y)+O(\alpha).$

At NLO, the photon can be emitted from either lepton produced by the electromagnetic current. For massless leptons, the polarization-averaged squared amplitude and the differential rate normalized to the tree-level cross-section are
\begin{equation}
\overline{|\mathcal M_3|^2}=\frac{16}{3}e^2 W~,\qquad
\overline{|\mathcal M_3|^2}\frac{d\Phi_3}{\sigma_0}=\frac\alpha\pi\,W\,dx_1dx_2\,,\qquad W=\frac{x_1^2+x_2^2}{2(1-x_1)(1-x_2)}\,,\qquad x_i=\frac{2E_i}{Q}\,.
\end{equation}
Here $x_1$ and $x_2$ are the lepton energy fractions, and $x_3 = 2-x_1-x_2$ is the photon energy fraction. 
Evaluating~\eqref{eq:sigma} for the three-body real emission process using
\begin{equation}
y=\frac{x_1+x_2-1}{x_1x_2}
\end{equation}
gives
\begin{equation}
\mathcal P(y)
=\int_0^1dx_1\int_{1-x_1}^1dx_2\,
W(x_1,x_2)\,
\delta\left(y-\frac{x_1+x_2-1}{x_1x_2}\right)
=\frac{-\log(1-y)-y-\tfrac12y^2}{y^3(1-y)}.
\end{equation}
Taking into account the normalization, leads to an additional subtraction of the Born weight, turning this into $[\cP(y)]_+$. Since virtual contributions cancel, the full quantity $\mathcal{D}(y)$ is given by ~\eqref{eq:Dist} to NLO.

To compute the two-point function~\eqref{eq:Oalpha} to NLO order we use Equation~\eqref{eq:final2pt} together with
\begin{equation}
    \int^{1}_v dy [\mathcal{P}(y)]_+=-\int^{v}_0 dy [\mathcal{P}(y)]_+=-\int^{v}_0 dy \,\mathcal{P}(y)=-\mathcal{R}(v),
\end{equation}
reproducing the NLO result~\eqref{eq:Oalpha}. The last integral is given by
\begin{equation}
\cR(s)=\frac54+\frac{1}{2s}+\frac{1+2s}{2s^2}\log(1-s)+\Li(s)+\frac12\log^2(1-s)\,.
\end{equation}
The corrections to the respective correlators are
\begin{align}
\delta_{zz}=\frac{\alpha}{\pi}\int_0^s\frac{dy}{1-y}\,\cR(y)
=\frac{\alpha}{\pi}\Bigg[\frac16L^3-\frac34L^2+L\left(\frac1{2s}+\frac54+\frac{\pi^2}{6}
+\operatorname{Li}_2(1-s)\right)-\frac32\operatorname{Li}_2(s)
+2\operatorname{Li}_3(1-s)
-2\zeta(3)-\frac12
\Bigg]
\end{align}
and
\begin{align}
\delta_{z\bar{z}}=\frac{\alpha}{\pi}\int_s^1\frac{dy}{y}\,\cR(y)
=\frac{\alpha}{\pi}\Bigg[\operatorname{Li}_3(1-s)-\operatorname{Li}_3(s)
+\zeta(3)+\frac{1-s}{4s}+L\left(\operatorname{Li}_2(1-s)
+\frac54-\frac1s-\frac1{4s^2}\right)
-\frac12L^2\log s
\Bigg]
\end{align}
with $L=-\ln(1-s)$.

\end{document}